# Preemptive punishment and the emergence of sanctioning institutions

Tatsuya Sasaki[1,2], Satoshi Uchida [3] and Xiaojie Chen[4]

[1] Faculty of Mathematics, University of Vienna, Vienna, Austria
[2] Department of Community Development, Koriyama Women's College, Koriyama-city, Fukushima, Japan
[3] Research Center for Ethi-Culture Studies, RINRI Institute, Tokyo, Japan
[4] School of Mathematical Sciences, University of Electronic Science and Technology of China, Chengdu, China
t.sasaki@koriyama-kgc.ac.jp

## Abstract

In explaining altruistic cooperation and punishment, the challenging riddle is how transcendental rules can emerge within the empirical world. Game-theoretical studies showed that pool punishment, particularly second-order punishment, plays a key role in understanding the evolution of cooperation. Second-order pool punishment, however, is tautological in nature; the punishment system itself is caused by its own effects. The emergence of pool punishment poses a logical conundrum that, to date, has been overlooked in the study of the evolution of social norms and institutions. Here, we tackle the issue by considering the interplay of (a) cognitive biases in reasoning and (b) Agamben's notion of *homo sacer* (Agamben, G. 1998, *Homo Sacer: Sovereign Power and Bare Life*, Stanford University Press); that is, a person who may be killed without legal consequences. Based on the cognitive disposition of reversing the cause-and-effect relationship, we propose a new system: the preemptive punishment of *homo sacer*. This action can lead to retrospectively forming moral assessments, particularly for second-order pool punishment. This study potentially provides new insight into the evolutionary process behind the emergence of social norms and institutions.

## Introduction

Cooperation in collective actions, such as extending mutual aid or providing public goods, poses a challenging conundrum (Olson, 1965) that has attracted broad attention from various scientific disciplines, from biology to the social sciences. Cooperation in collective action tends to be costly, and this, in turn, gives rise to the temptation to freeload on the contributions of others (the so-called "free rider" problem). This will lead to terminating voluntary cooperation among rational individuals or under Darwinian dynamics, unless other factors or mechanisms of support are involved.

The prescription of selective incentives for encouraging contributors and/or deterring free riders is one of the most studied resolutions of the free rider problem (Olson, 1965; Oliver, 1980; Coleman, 1988). Clearly, it is the provision of selective incentives that itself tends to be costly and thus constitutes another public good. Those who freeload others' contributions to the incentive system are likely to emerge. This so-called "second order" free rider problem has already been tackled by interdisciplinary studies (Oliver, 1980; Heckathorn, 1989; Yamagishi, 1986; Horne, 2001).

To date, game-theoretical studies on the evolution of cooperation have demonstrated that differences in the details of punishment systems can have a large effect on the evolutionary fate of human cooperation (Boyd and Richerson, 1992; Sigmund et al., 2010; Sasaki et al., 2012). In particular, the main differences are (a) whether the decision to punish is reactive or proactive and (b) whether the punishment of second-order free riders is considered or not.

## Peer and pool punishments

One representative type of punishment, peer punishment, is often inductively modeled, typically described as, "Because you wronged me (or someone), I will punish you." The evolution of peer punishment thus depends on initial conditions and an assessment of past behaviors and can be studied in line with direct reciprocity for iterated interactions (Trivers, 1972) or indirect reciprocity for social networks with reputation or gossip media (Alexander, 1987). A voluntary punisher who imposes penalties on free riders extends a public good that contributes to relatively increasing the utility or fitness of all others. Considering that peer punishment is likely to incur risks of retaliation or counter-punishment, such costly peer punishment will often lead to tempting people to free ride on others' punishing efforts. This can pave the way for regression to the second-order punishment of free riders through peer punishers. The same logic applies to third-order punishment (Colman, 2006).

Even if the population has achieved 100% cooperation through peer punishment, another problem occurs. The absence of first-order free riders will result in difficulties discriminating punishers (second-order contributors) from non-punishers (second-order free riders). This then leads to no selection pressure, and thus, a neutral drift between those second-order actions. By neutral drift, the population can lose a substantial fraction of the punishers. This means that mutant first-order free riders will be able to invade the population more easily.

Ironically, the goal of establishing cooperation can be well maintained by not completely eliminating free riders. Without the appropriate management of mutant free riders, a cascade of collapses of punishment systems may occur. As the aforementioned regression develops, the same logic applies between third-order contributors and free riders. The neutral drift breaks if both first-order and second-order free riders are present.

As the group size grows, these problems may lead to participants' convictions that punishments are continuously necessary, that free riders are supposed to be somewhere in every interaction, and thus, it becomes impossible to achieve an ideal state of 100% cooperators.

Another representative type of punishment is pool punishment (Yamagishi, 1986; Sigmund et al., 2010). The type of pool punishment expected here is a prospective scheme that can be described as, "Free riders (no matter for what public good) are supposed to be somewhere, and thus the system needs to punish free riders of public goods." This is followed by, "Thus, one ought to contribute to such comprehensive punishment, or otherwise one would themselves be similarly punished." In its standard model of public good games (PGGs), pool punishment is set in place before the establishment of the PGG, and each participant is then offered the opportunity to contribute to a fund for establishing the pool punishment system (Sigmund et al., 2010). It is also assumed that pool punishment becomes active if at least one player contributes to the fund; otherwise, the system will not be implemented because of a lack of funds.

Game-theoretical studies showed that when considering the punishment of second-order free riders for those who fail to contribute to the fund, pool punishment becomes more effective than peer punishment in stabilizing a cooperative state, and participants are likely to prefer pool punishment over peer punishment (Sigmund et al., 2010; Perc, 2012; Traulsen et al., 2012; Hilbe et al., 2014).

## Second-order pool punishment

The essence of pool punishment is its prior commitment system and second-order punishment module. As long as the participants commit to pool punishment, it is not difficult to install a prospective mechanism, such as taking deposits and hostages and hiring a sheriff, to implement 'first-order' punishment of free riders for the PGG. First-order punishment is based on the first-cause-then-effect policy that an effect cannot occur before its cause and can naturally be given legitimacy by traditional reciprocal justice.

However, here we do not assume a prior social norm for legitimizing the second-order punishment of free riders for the punishment fund. That is, in the first place, a conscious pool punisher is assumed to be present, but it is not that there is common knowledge of an assessment rule for judging non-contribution to pool punishment. As such, it would not be easy to lead participants to commit to the specific norm of pool punishment in which second-order free riding is assessed as bad.

This is to say that each participant is not supposed to transcendentally abide by the assessment rule. Therefore, when a participant makes a decision regarding contribution on the basis of a moral assessment by the system, its attempt seems to be tautology, because the moral assessment rule, when once committed by the participant, is applicable to itself. Therefore, it is implied that pool punishment must have originated empirically, while almost all individuals are likely to recognize pool punishment as if it has already been given *a priori*.

## Symmetry in thinking

Here, we recall the moral assessment considered in indirect reciprocity (Sugden, 1986). Indirect reciprocity through reputation is often described by using the helper's reputation as follows: "If I have helped you, then I will have a good image, and then someone will help me" (Kandori, 1992; Nowak and Sigmund, 1998; Ohtsuki and Iwasa, 2004; Brandt and Sigmund, 2004; Uchida and Sigmund, 2010). Although models of indirect reciprocity mostly lack the dynamics to address the recipient's reputation, updating the recipient's reputation often happens in daily life. Assuming that "If you are bad, you will be punished" is true, this is described through the following reasoning: "You must have been bad, because someone has punished you, and then you do not pay it back." (That is, "You have been responsible (namely, 'guilty') for something that caused the punishment.")

This is the so-called fallacy of affirming the consequent, and such cognitive disposition of reversing the cause-and-effect relationship is called "symmetry bias," a heuristic way to overcome the trade-off dilemma of exploration and exploitation of information (Nakano and Shinohara, 2008). This topic has been explored in a great deal of studies since Kahneman and Tversky's work (Kahneman and Tversky, 1973). For instance, if probabilities $P(A)$ and $P(B)$ of events A and B are given together with a conditional probability $P(A|B)$, the reverse conditional probability $P(B|A)$ must be calculated by Bayes' theorem as $P(B|A) = P(A|B)P(B)/P(A)$. However, human beings tend to omit the base rate $P(B)/P(A)$ in estimating the value of the conditional probability $P(B|A)$ without following Bayes' theorem. As a result, they reach the wrong guess that $P(B|A) = P(A|B)$ (Bar-Hillel, 1980).

Symmetry bias has also been underlying our economy. Originally, commodity C plays the role of money in a society because the values of other commodities in the society are measured contingently by the amount of C. However, people gradually think that commodity C intrinsically has the characteristic of money (thus, other commodities can never play the role of money), and that therefore the values of other commodities are measured by the amount of C. This inversion of logic in the economic context is referred to as the "commodity fetishism" of money by Karl Marx in the first chapter of his famous book *Capital: Critique of Political Economy* (Marx, 1909).

We assume that a similar cognitive error occurs in the context of indirect reciprocity. Previous models of indirect reciprocity through reputation have mostly assumed that a (bad or unknown) player is able to affect its reputation only after its own action. For our model of pool punishment, we take symmetry bias into consideration, which means that a punisher's actions can affect a punishee's reputation and even make it up out of nothing.

## *Homo sacer* and preemptive punishment

Taking into account the effects of symmetry bias on moral assessment, therefore, one of the remaining missing pieces would be to determine who is an appropriate target to be punished for inventing and legitimizing a moral rule when in the presence of the whole group ("third party"). Referring to Masachi Ohsawa's introduction and interpretation (Ohsawa, 2003), let us apply Giorgio Agamben's concept of a person called *homo sacer* (Agamben, 1998) to our model. *Homo sacer* (a Roman legal term transferred as "sacred man") is described as someone excluded from the law itself, and anybody who has killed this person is acquitted of the charge for the killing. That is, *homo sacer* is an outlaw in its literal sense, like irregular migrants and stateless persons in modern societies. At the beginning of joint efforts, the innovative punisher, who is willing to continuously police *would-be* free riders, should actively punish a member like a *homo sacer*. From the viewpoint of the law of fitness, changing things or even something transitory, a feature of *homo sacer*, seems to prefer more profitable things and thus at least does not commit to costly punishment.

For this reason, the nature of *homo sacer* will work as a trigger of its preemptive sanction from the strictest and most costly continuous punisher (X in Fig. 1). To a third party (Z in Fig. 1), there is no ground for understanding badness other than the fact of *homo sacer*'s change and non-contribution. Considering the aforementioned symmetric bias, this can lead to faulty reasoning: "The target (Y in Fig. 1) deserves being punished, because its nature has been bad," and "Possibly, therefore, changing (from "contribution") to non-contribution has been assessed as bad." This may result in another faulty reasoning (the so-called mutual exclusivity bias): "Contribution has been assessed as good, and from the beginning it ought to be that a man does not alter his contribution."

Of course, the cause of punishment of *homo sacer* is nonsense, and s/he is innocent of anything (and thus, in particular, of free riding), because initially there has been no norm for the validity of non-contributing to the pool punishment and everyone has not acted (and thus paid) for it. To perfectly project the transcendental cognition of norms by the preemptive sanction of *homo sacer*, the existence of *homo sacer* should subsequently be eliminated. This means that *homo sacer* can be excluded twice, because the nature of *homo sacer* has already been outside of both the law and courts (Ohsawa, 2003).

Thus, the ban on *homo sacer* leads to the completion of the system's consistency in pool punishment, that is, with a definition of goodness and badness. By rejecting *homo sacer*, which is a point of inconsistency, the resulting assessment rule can be viewed as transcendental. Thus, the larger the number of individuals who are jointly included in attention to ceremonial preemptive punishment, the more the assessment rule may be spread over the population by support of cognitive bias. This could resolve the coordination problem of pool punishment.

## Quicksand and scarecrow

From what we discuss in this paper, one may recall a puzzle of rules and communication: forming consensus of a rule itself needs consensus of another rule in advance. How can infinite regression stop? How can the initial consensus of knowledge be coordinated? To better understand the implications of the puzzle for our model, we refer to the example of the quicksand and the scarecrow by the analytic philosopher David Lewis (Lewis, 1969). With respect to the example, Bardsley and Sugden (2006, pp. 763-764) write:

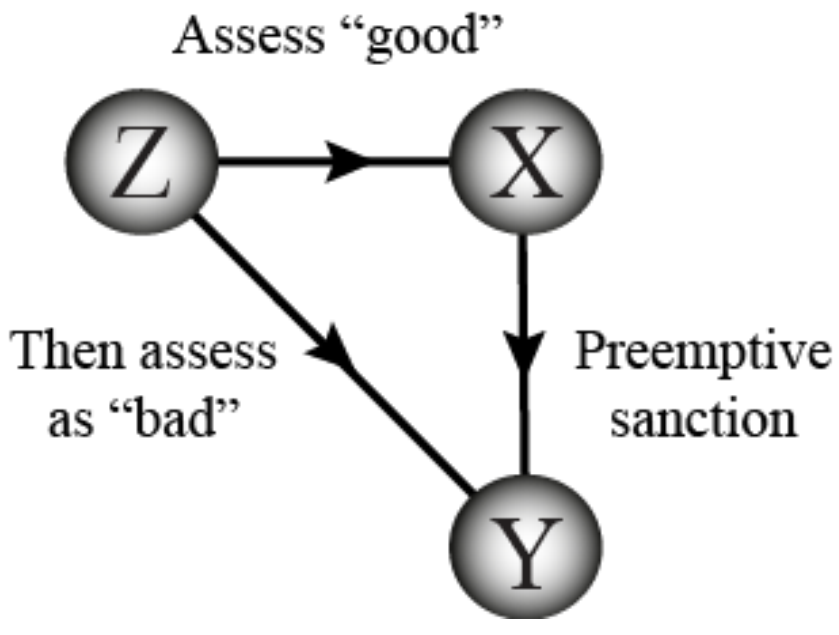

Fig.1. Establishing moral assessment rules through preemptive sanctions. Player X sanctions player Y so that the resulting assessment rule leads observer Z to retrospectively assess punisher X as good and punishee Y as bad. If punishee Y waives its claim of counter-punishment of punisher X, then the retrospective system based on preemptive sanctions can be justified.

> Lewis (1969, p. 158) imagines coming across a patch of quicksand, waiting to warn others of the danger, but not knowing of any existing conventional signal. So: 'I put a scarecrow up to its chest in the quicksand, hoping that whoever sees it will catch on.' Although this signal does not yet have any conventional meaning, Lewis says, 'I have done my part of a signaling system in a signaling problem, and I hope my future audience will do its part.' This example seems to establish that an agent could *mean something by* his placement of the scarecrow without its having a prior use in any community. This meaning seems to be established by the agent's intention and appears to be responsible for its later taking on a conventional meaning, that half-submerged scarecrows *stand for* quicksand.

Considering Lewis's images in line with our model, it looks like the person who warned others of danger corresponds to the innovative punisher (who initially exemplifies punishment prior to joint efforts). The *homo-sacer*-like punishee who suggests the existence of a moral rule could be viewed as the half-submerged scarecrow that stands for quicksand. This is because the *homo sacer*, who has less identity, is extremely passive in asserting its own right, like the scarecrow that was at the mercy of the signaler. In the case of quicksand, the scarecrow may have been possessed by the signaler or like a stone on the wayside that anybody may pick up and thus could be disposed of into the quicksand as the signaler liked. Similarly to such a stone, *homo sacer* seems to be a private and nighest thing.

Let us return to what we discussed about the repeated regression of peer punishment in the early section. From the nature of peer punishment, it is understood that the higher the order of the punisher, the more difficult it is to discriminate the punisher from those who almost always contribute to public goods, but free ride only on the last order of punishment.

Therefore, the degree of ambiguity of higher-order free riders will reach the limit when considering the infinite regression of peer punishment. There is only a very fine line between the resulting punisher and the punishee. In this case, both are able to transform into each other by only very minor changes. This implies that the continuous peer punisher could almost be itself its punishee, who is viewed as a *homo sacer* for the preemptive sanction. That is, the system of peer punishment is rejected by itself within the limits of regression. However, this failure means nothing other than a path to success in establishing a moral assessment for pool punishment.

## Conclusion

Our discussion has been extended from featuring the characteristics of peer and pool punishments to exploring how the transition between those punishment can emerge. Our conjecture is that the creation of a moral assessment rule for pool punishment needs to exclude contingency or transitoriness, often described as mutation or exploration which is one of the fundamental principles of evolution. The discussion provides new insight into the evolutionary process underlying the establishment of sanctioning institutions.

Those who survive in a competitive world governed by the law of fitness should have something transitory to a greater or lesser degree. In the sense, it is maybe that we ourselves could be *homo sacer*, which can represent uncertainty in our evolving lives, and that for making moral judgment, we need the option of cutting off the changeable *homo-sacer* part from the whole. This will leave behind the other, unchangeable “zealous” part, which is not interested in maximizing fitness (Masuda, 2010) and is capable of controlling our normative life by means of pool punishment.

By excluding evolvability, such as mutation and exploration, a cooperative punisher can opt out of the Darwinian dynamics, the competitive games of life. That is, the zealous punisher will be viewed as a kind of environmental parameter to stabilize sanctions in the evolutionary game. This may be described as the “second-order” optional participation, which is a meta level of the original, first-order optional participation: to opt out of particular joint efforts, while instead relying on a small payoff independent of what others do (Hauert et al., 2002). To better model and analyze the evolution of social norms, therefore, it would be fascinating to collaborate with previous studies on evolving mutation rates (Kaneko and Ikegami, 1992) and excludable public good games (Sasaki and Uchida, 2013; Liu et al., 2009).

## Acknowledgement

We thank Karl Sigmund. T.S. was supported by the Foundational Questions in Evolutionary Biology Fund: grant no. RFP-12-21 and the Austrian Science Fund (FWF): P27018.